# MPI 程序同步通信基本模型死锁检测

廖名学，范植华

(中国科学院软件研究所, 北京 100080)

**摘 要**：本文提出了MPI程序的同步通信模型及三个基本简化模型，给出了判定这些基本模型是否死锁的方法和定理并予以了严格证明. 简化模型的死锁检测理论和方法是真实MPI程序死锁检测的必要基础. 这些方法基于程序静态分析，必要时进行运行时检测，它们对两种简化模型可以在程序编译前确定是否死锁，对另外一种模型，在编译前可静态确定部分死锁，运行中可确定其他死锁. 我们的理论可以证明MPI程序死锁检测主流算法的正确性，其方法可以减少它们对客户源代码或MPI profiling接口的修改量，从而大大降低死锁检测开销，并可在运行前判定死锁.

关键词：MPI(message passing interface)；死锁；同步通信

## Deadlock Detection in Basic Models of MPI Synchronization Communication Programs

LIAO Ming-xue, FAN Zhi-hua

(Institute of Software, the Chinese Academy of Sciences, Beijing 100080, China)

**Abstract**：A model of MPI synchronization communication programs is presented and its three basic simplified models are also defined. A series of theorems and methods for deciding whether deadlocks will occur among the three models are given and proved strictly. These theories and methods for simple models' deadlock detection are the necessary base for real MPI program deadlock detection. The methods are based on a static analysis through programs and with runtime detection in necessary cases and they are able to determine before compiling whether it will be deadlocked for two of the three basic models. For another model, some deadlock cases can be found before compiling and others at runtime. Our theorems can be used to prove the correctness of currently popular MPI program deadlock detection algorithms. Our methods may decrease codes that those algorithms need to change to MPI source or profiling interface and may detects deadlocks ahead of program execution, thus the overheads can be reduced greatly.

**Key words**：message passing interface (MPI); deadlock; synchronization communication

## 1 引言

死锁问题由来已久. 1968 年Dijkstra描述了有限资源环境中并发进程资源请求过程可能出现的死锁并给出了银行家算法. 1971 年Coffman给出了死锁产生的 4 个必要条件以及处理死锁的三种策略：死锁预防，死锁避免与死锁检测和恢复. 文**错误！未找到引用源。**认为死锁预防和死锁避免具有很多缺点，难度很高，主要用在可靠性要求甚高的系统中，因此Mukesh[2]认为死锁处理大多采用死锁检测. 死锁检测在20世纪中在数据库和分布式系统方面. 其中，分布式系统死锁检测方法一般分为集中式（使用处理控制中心），分布式（难点在于维护全局状态图）和层次式（站点采用树形组织）. 近几年则在死锁预防和死锁检测上取

得了很多带有鲜明具体程序语言特征的成果.Ryder[3]将它们分为了流敏感，上下文敏感，域敏感，程序表示，对象表示和引用表示等几类（这几类并不是绝对隔离的，在某种具体方法中常常是结合使用的）.Christoph von Praun[4]给出了一种多线程面向对象程序同步死锁检测算法.文[5]给出了JDK中死锁检测算法.Tong[6]提出了一个运用于Pulse操作系统机制的称为推测执行的动态死锁检测方法，其需要修改和增加内核.RacerX[7]是一种用于C系统代码的流敏感上下文敏感的死锁和数据竞争检测工具，它需要对系统特定函数中加锁行为进行注解.Breuer和Valls[8]描述了Linux内核中的静态死锁检测——检测调用sleep且拥有spinlock的线程产生的死锁.

    MPI[9]是分布式内存并行处理计算机上开发基于消息传递应用系统的标准（其主要实现是MPICH[10]），主要用于大规模并行计算机和集群的高性能运算，但其调试极其困难[11]..目前MPI上的调试很大程度上集中于通信死锁检测，一般都采用动态方法（运行时），目前有四种主流方法和工具. 文[12]的方法需要在原程序代码中所有MPI调用语句之前插入握手代码收集各节点状态来判断是否死锁.文[11]则利用MPI Profiling接口，截获用户的所有MPI调用并进行死锁分析.文[13]类似[11]，但需要在共享内存架构上执行MPI程序.文[14]统计MPI程序中的各种错误，也采用MPI Profiling接口跟踪MPI调用，只是检测范围不仅仅只包括死锁.采用静态方法检测死锁的形式化方法和工具比较典型的是SPIN[15]，它首先要根据具体的程序给出模型规格说明和属性定义.

## 2 MPI 程序同步通信模型及其基本模型

### 2.1 MPI 程序同步通信模型描述

    根据Jacopini[16]中阐明的程序结构定理，若只考虑MPI程序中与同步通信相关的代码，那么MPI程序同步通信模型可用图 1 表示出来：

---

Program → node_program+ /*MPI 程序由在各个结点机上运行的程序(结点程序)构成*/

node_program → <nodeID,statements>/*结点程序由结点标识和运行语句集(语段)构成*/

statements → statement*/*语段由若干条语句构成*/

statement → sequence_statement|if_statement|for_statement/*语句分顺序/条件/循环三种*/

sequence_statement → (send|receive)+/*顺序语句是若干 MPI 同步通信发送/接收语句*/

if_statement → if(condition){statements}/*条件语句由条件和语段构成*/

for_statement → for(i:i1→i2){statements}/*循环语句由循环变量上下界和语段构成*/

---

图 1   MPI 程序同步通信模型

这种模型称为 $M_0$.从 MPI 标准来看，$M_0$ 仅仅涵盖其中的点对点同步通信.$M_0$ 又可细分为若干基本子模型.仅含顺序语句的 $M_0$ 称为顺序模型 $S_0$.仅含一条条件语句且该条件语句的语段都是顺序语句的 $M_0$ 称为条件模型 $C_0$.仅含一条循环语句且该循环语句的语段都是顺序语句的 $M_0$ 称为循环模型 $L_0$.

## 2.2 相关定义

**[定义1]  消息与宿主**

消息（同步收发语句）$m = \{from, method, contents, to\}, method \in \{send, recv\}$.宿主是 MPI 并行计算的一个节点，用正整数表示.$from$ 表示消息源宿主，$to$ 表示消息目的宿主，$contents$ 表示消息内容，$method$ 表示消息处理方式：发送或接收.$m.from$ 表示 $m$ 的消息源宿主，其余类推.

**[定义2]  序列**

$s = x_1 x_2 ... x_n$ 称为序列，若 $x_i$ 为序列或终结符，记 $x_i \in s$.序列可看作有序集合.

**[定义3]  简单消息序列（集）与消息（序列）宿主函数**

$s = m_1 m_2 ... m_n$ 称为简单消息序列，若 $m_i$ 是消息.消息 $m$ 宿主函数定义为 $host(m) = \begin{cases} m.to \ iff \ m.method = recv \\ m.from \ iff \ m.method = send \end{cases}$. $s$ 宿主函数定义为 $host(s) = host(m), m \in s$，即 $s$ 中出现的所有消息的宿主都是一样的，也即消息序列对应一个节点计算机上执行的程序（不考虑非 MPI 调用语句）.消息序列的有限集称为消息序列集.

**[定义4]  消息序列集死锁和死锁函数**

消息序列集 $L$ 死锁是指 $L$ 中存在一条消息，该消息的接收或者发送动作已经做出，但它或者永远接收不到消息内容或者消息内容永远发不出去.死锁函数 $deadlock(L)$ 当 $L$ 死锁时值为逻辑真 $T$，否则为逻辑假 $F$.

## 3   顺序模型死锁分析

### 3.1 顺序模型相关定义

**[定义5]  $S_0$ 型序列（集）与序列长度函数**

简单消息序列 $s$ 称为 $S_0$ 型序列；$S_0$ 型序列的有限集合称为 $S_0$ 型序列集.$s$ 的长度函数 $length$ 定义为 $s$ 中消息总数，用 $s.length$ 或 $length(s)$ 表示.

**[定义6]  $S_0$ 型序列消息先后关系**

设 $<$ 是 $S_0$ 型序列 $s = m_1 m_2 ... m_n$ 上的关系，且 $<= \{< m_i, m_j > | 1 \le i < j \le n\}$，则 $<$ 称为 $s$ 上的消息

先后关系，表示消息 $m_i$ 先于 $m_j$ 发生（记作 $m_i < m_j$）. $<$ 是反自反，反对称和传递关系.

**[定义7]　$S_0$ 型序列集消息匹配关系与全匹配**

设 $match$ 是 $S_0$ 型序列集 $A = \{s_i \mid 1 \le i \le n\}$ 上 的 关 系，

$match = \{<m_p, m_q> \mid m_p \in s_i \wedge m_q \in s_j \wedge m_p.from = m_q.from \wedge m_p.to = m_q.to \wedge m_q.method \ne m_p.method\}$ 且

$(\neg\exists m_x)((m_x < m_p \wedge (m_x, m_q) \in match) \vee (m_x < m_q \wedge (m_p, m_x) \in match))$，那么 $match$ 称为 $A$ 上的匹配关

系. $<m_i, m_j> \in match$ 可记作 $match(m_i, m_j)$. 匹配关系实质上就是消息的发送/接收关系.由于消息

是消耗性资源[17]，从而一条消息匹配多条消息没有实际意义，因此如果消息 $m_1$ 与某条消息

$m_2$ 匹配，就不可能与在 $m_2$ 后面发生的消息匹配. $match$ 是反自反，对称和非传递关系.

若 $S_0$ 型序列集 $A = \{s_i \mid 1 \le i \le n\}$ 满足 $(\forall m_p \in s_i)(\exists m_q \in s_j)(<m_p, m_q> \in match)$ 那么 $A$ 是全匹配的，

否则 $A$ 是非全匹配的. 即，若有某个消息无匹配则 $A$ 是非全匹配的，否则是全匹配的.

**[定义8]　$S_0$ 型序列集消息同序关系**

$S_0$ 型序列集 $A = \{s_i \mid 1 \le i \le n\}$ 上消息恒等关系与匹配关系的并集称为消息同序关系，记为

$=$ . $<m_i, m_j> \in =$ 可记作 $m_i = m_j$，表示消息 $m_i$ 和 $m_j$ 或者是同一条消息（注意消息序列是有序集

合），或者是互相匹配的消息. $=$ 是等价关系.

**[定义9]　$S_0$ 型序列集消息非迟于关系**

$S_0$ 型序列集上的消息同序关系与消息先后关系的并称为 $A$ 上的非迟于关系，记为 $\le$ .若

$m_i \le m_j$，则消息 $m_i$ 不迟于 $m_j$ 发生. $\le$ 自反，反对称，传递.

**[定义10]　$S_0$ 型序列集依赖图**

给定 $S_0$ 型序列集 $A = \{s_i \mid 1 \le i \le n\}$ 与空图 $G$. 将消息看作点加入到 $G$. 对消息 $m_i, m_j$，若

$match(m_i, m_j)$ 则增加双向边 $m_i \leftrightarrow m_j$；若同序列且 $j - i = 1$ 则增加有向边 $m_i \to m_j$. 即构成依赖图.

**[定义11] $S_0$ 型序列集关联关系与关联划分**

$D\_relation = \{<s_i, s_j> \mid s_i, s_j \in A, (\exists m_p \in s_i, m_q \in s_j)(<m_p, m_q> \in match)\}$ 称 为 $S_0$ 型 序 列 集

$A = \{s_i \mid 1 \le i \le n\}$ 上 的 直 接 关 联 关 系，表 示 两 个 序 列 之 间 存 在 一 个 消 息 匹 配，记 为

$<s_i, s_j> \in D\_relation$ .称 $relation = t(D\_relation) \cup I_A$（ $D\_relation$ 传递闭包与 $A$ 上恒等关系的并集）

为关联关系，记为 $relation(s_i, s_j)$ . $relation$ 是等价关系；$D\_relation$ 是对称，反自反，非传递关

系.$relation$ 所决定的划分称为 $A$ 之关联划分.

**[定义12] $S_0$ 型序列（集）加运算**

同宿主的两个 $S_0$ 型序列的 "$+$" 运算就是连接运算(类似字符串连接)；不同宿主的 $S_0$ 型序列的加运算结果为空集. $S_0$ 型序列集 $A$ 和 $B$ 定义为 $A + B = \{a+b | a \in A, b \in B\}$.

**[函数1] $S_0$ 型序列前缀函数和后缀函数**

$S_0$ 型序列 $s = m_1 m_2 ... m_n$ 前缀函数 $prefix(s,p)$ 定义为 $s$ 中最先发生的 $p$ 条消息；后缀函数 $suffix(s,p)$ 定义为 $s$ 中最后发生的 $p$ 条消息.

**[定义13] $S_0$ 型序列（集）减运算**

同宿主的两个 $S_0$ 型序列 $s_1, s_2$ 的右减 "$-^R$" 运算 $s_1 -^R s_2$ 是将 $s_1$ 的后缀 $s_2$ 删除(假定 $s_1$ 的后缀为 $s_2$，否则该运算结果为空集)；左减 "$-^L$" 运算 $s_1 -^L s_2$ 是将 $s_1$ 的前缀 $s_2$ 删除(假定 $s_1$ 的前缀为 $s_2$，否则该运算结果为空集).不同宿主的 $S_0$ 型序列的右减和左减运算结果为空集. $S_0$ 型序列集 $A$ 和 $B$ 的左减运算定义为 $A -^L B = \{a -^L b | a \in A, b \in B\}$，右减运算可类推.

## 3.2 顺序模型的重要性质

**[性质1] $S_0$ 型序列集依赖图性质**

$S_0$ 型序列集 $A$ 的依赖图 $G$ 与 $A$ 上的匹配关系和先后关系的并集是等价的.

证明：(1)根据[定义 10]，$G$ 点集与 $A$ 中出现的消息一一对应.

(2)根据[定义 10]，$G$ 的单向边集合是 $A$ 的先后关系的子集，双向边集合就是 $A$ 的匹配关系. $A$ 上的先后关系是传递的，该关系的任何一个元素都可以在 $G$ 上找到一条路与之对应.

**[性质2] $S_0$ 型序列集的可加性和可减性**

设有 $S_0$ 型序列集 $A = \{s_i | 1 \leq i \leq n\}$ 及其上的一对匹配消息 $match(m_1, m_2)$，则有：

$deadlock(A + \{\{m_1\}, \{m_2\}\}) = deadlock(A)$，

$deadlock(A -^R \{\{m_1\}, \{m_2\}\}) = deadlock(A)$ 与 $deadlock(A -^L \{\{m_1\}, \{m_2\}\}) = deadlock(A)$.

证明：只证 $deadlock(A + \{\{m_1\}, \{m_2\}\}) = deadlock(A)$.

(1)若 $A$ 死锁，即在 $m_1, m_2$ 之前有 $k > 0$ 个消息处于永远等待状态.只需要考虑 $k = 2$ 情形. 设这两个消息为 $m_1', m_2'$. 若 $host(m_1') = host(m_2')$，但 $m_1, m_2$ 宿主不同，死锁不能解除. 若 $host(m_1') \neq host(m_2')$，若两条消息非所在序列的最后一条消息或 $m_1', m_2'$ 不能与 $m_1, m_2$ 分别匹配，则死锁不能解除；若分别匹配，不妨设 $match(m_1, m_1'), match(m_2, m_2')$，故 $m_1, m_2'$ 同宿主，$m_2, m_1'$ 同宿主，此时形成循环等待，仍然死锁.

(2)若 $A$ 无死锁,即在 $m_1, m_2$ 之前没有任何消息处于等待状态.在 $A$ 之后增加 $m_1, m_2$,待 $A$ 中所有消息完成之后,剩下的消息 $m_1, m_2$ 也可以执行完毕.故不会引入新死锁.

**[性质3]  $S_0$ 型序列集可加性可减性推论**

$S_0$ 型序列集 $A, B$ 满足: $deadlock(A) = deadlock(A + B)$ $if$ $deadlock(B) = F$ .（证明略）

$deadlock(A) = deadlock(A -^R B)$ $if$ $deadlock(B) = F$ .（证明略）

$deadlock(A) = deadlock(A -^L B)$ $if$ $deadlock(B) = F$ .（证明略）

### 3.3 顺序模型死锁判定定理

**[定理1]  $S_0$ 型序列集死锁必要条件**

$S_0$ 型序列集 $A = \{s_i \mid 1 \le i \le n\}$ 无死锁的必要条件是 $A$ 全匹配. （证明略）

**[定理2]  $S_0$ 型序列集死锁充要条件 1**

全匹配 $S_0$ 型序列集 $A = \{s_i \mid 1 \le i \le n\}$ 及其上的关联划分 $D$ ,则

$deadlock(A) = deadlock(d_1) \vee deadlock(d_2) \vee ... \vee deadlock(d_{|D|})$ $d_i \in D$ .（证明略）

**[定理3]  $S_0$ 型序列集死锁充要条件 2**

全匹配 $S_0$ 型序列集 $A = \{s_i \mid 1 \le i \le n\}$ 及其上的依赖图 $G$ . $A$ 的死锁性等价于 $G$ 中有无长度大于 2 的圈[18].即若 $G$ 有长度大于 2 的圈,则 $A$ 死锁;否则, $A$ 无死锁.

证明：根据[性质1]知 $A$ 与 $G$ 一一对应.

(1)若 $G$ 有长度大于 2 的圈,则 $A$ 死锁.

设长度大于 2 的圈为 $<v_1, v_2, ..., v_m, v_1>$ ,由于 $G$ 的顶点与消息对应; $G$ 的边与消息先后关系和匹配关系对应. 故有: $v_1 \le v_2 \le ... \le v_m \le v_1$ ,根据传递性,可知 $v_1 \le v_1$ ,并仅当 $v_1 = v_2 = ... = v_m = v_1$ 时 $v_1 = v_1$ 成立,否则 $v_1 < v_1$ .假设 $v_1 = v_2 = ... = v_m = v_1$ ,由于圈长度大于 2,可知至少有两个点与 $v_1$ 等价,从而有两条消息与 $v_1$ 对应的消息匹配,这是不可能的.从而必有 $v_1 < v_1$ .故 $A$ 死锁.

(2)若 $G$ 无长度大于 2 的圈,则 $A$ 无死锁.

假定 $A$ 死锁,则必存在同一个消息序列的两点 $v_1 < v_2$ ,使得 $v_2 < v_1$ 也成立,但 $v_2$ 到 $v_1$ 的路径必定经过了不同序列的某个点,如此形成的圈长度必定大于 2.

**[例1]  $S_0$ 型序列集死锁分析举例[消息采用简化记法]**

表 1 顺序模型死锁分析举例

| $s_1$ | $s_2$ | $s_3$ | G |
|---|---|---|---|
| send(a) | recv(b) | recv(c) |  |
| send(b) | send(c) | recv(c) | |

| 依赖图含长度 6 的圈，必死锁 | |
|---|---|

## 4 循环模型死锁分析

### 4.1 循环模型的相关定义

#### [定义14] $L_0$ 型序列（集）及其顺序式

序列 $s = Ns'$ 称为 $L_0$ 型序列，其中 $N > 0$ 表示循环次数，用 $s.N$ 表示；$s'$ 是循环体中的 $S_0$ 型序列，用 $s.Sequence$ 表示，称为 $s$ 的顺序式. $L_0$ 型序列的有限集合 $A = \{s_i \mid 1 \le i \le n\}$ 称为 $L_0$ 型序列集，其顺序式 $A.Sequence$ 为 $\cup s_i.Sequence$ .

#### [定义15] $L_0$ 型序列（集）展开式

对 $L_0$ 型序列 $s = Ns'$ ，下面的过程构造出其展开式 $s.Expand$ ：

$$s.Expand = \phi$$
$$\text{for}(1 \le i \le s.N)\ s.Expand = s.Expand + s.Sequence$$

显然，$s$ 与 $s.Expand$ 等价. $L_0$ 型序列集 $A = \{s_i \mid 1 \le i \le n\}$ 的展开式 $A.Expand$ 为 $\cup s_i.Expand$ .

#### [定义16] $L_0$ 型序列集匹配关系

设 $match$ 是 $L_0$ 型序列集 $A = \{s_i \mid 1 \le i \le n\}$ 上的关系且：

$match = \{< m_p, m_q > \mid m_p.from = m_q.from \wedge m_p.to = m_q.to \wedge m_p.method \ne m_q.method\}$

那么 $match$ 称为 $A$ 上的匹配关系.与 $S_0$ 型序列集匹配关系的区别在于，$L_0$ 型序列集匹配关系允许某个 $L_0$ 型序列中的一条消息匹配其他某个 $L_0$ 型序列中的多条消息.

#### [定义17] $L_0$ 型序列集消息周期比

设某个消息 $m$ 在某个 $L_0$ 型序列 $s_1$ 的顺序式中出现 $p(p \ge 1)$ 次，设与该消息匹配的消息在 $L_0$ 型序列 $s_2$ 的顺序式中出现 $q(q \ge 1)$ 次；称 $q : p$ 为消息 $m$ 在 $s_1$ 中的周期比，记为 $T(m, s_1) = q : p$ ；称 $p : q$ 为消息 $m$ 在 $s_2$ 中的周期比，记为 $T(m, s_2) = p : q$ .

#### [定义18] $L_0$ 型序列集序列比

$L_0$ 型序列集 $L$ 及其顺序式 $A = \{s_i \mid 1 \le i \le n\}$ . 若 $(\forall m_p, m_q \in s_i)(\forall m_x, m_y \in s_j)(match(m_p, m_x) \wedge match(m_q, m_y) \rightarrow T(m_p, s_i) = T(m_q, s_i))$ ，则 $T(m_p, s_i)$ 称为 $s_i, s_j$ 的序列比，简记为 $s_i : s_j$ ；否则，称 $s_i : s_j$ 不存在.

#### [定义19] $L_0$ 型序列集最简周期表

$L_0$ 型序列集顺序式的关联划分 $D$ .对任意 $d = \{s_i \mid 1 \le i \le m\} \in D$ ，若 $s_1 : s_2 : \ldots : s_m$ 存在，则其最简整数比称为 $d$ 最简周期表，记为 $T_1 : T_2 : \ldots : T_m$ ；否则，称 $d$ 最简周期表不存在.

### 4.2 循环模型死锁判定定理

**[定理4]** **$L_0$型序列集展开式全匹配必要条件**

$L_0$型序列集展开式全匹配必要条件是任意直接关联序列 $s_i, s_j$ 的序列比存在.

证明：设 $L_0$ 型序列集为 $A = \{s_i \mid 1 \le i \le n\}$，$D\_relate(s_i.Sequence, s_j.Sequence)$.假设 $s_i : s_j$ 不存在.

不妨设 $s_i$ 上不同消息 $m_1 m_2$ 在 $s_j$ 中有匹配消息，根据假设有 $T(m_1, s_i) = q_1 : p_1 \ne T(m_2, s_i) = q_2 : p_2$.则对任意正整数对 $(p, q)$，由于 $q_1 * q : p_1 * p \ne q_2 * q : p_2 * p$，故 $prefix(s_i.Expand, p * s_i.Sequence.length)$ 与 $prefix(s_j.Expand, q * s_j.Sequence.length)$ 中 $m_1, m_2$ 总有不能匹配的消息.

**[定理5]** **$L_0$型序列集无死锁必要条件**

$L_0$ 型序列集 $L$ 及其顺序式关联划分 $D$.$L$ 无死锁，则任意 $d \in D$ 最简周期表总存在.

证明：假设 $d$ 最简周期表不存在.记 $d = \{s_i \mid 1 \le i \le m\}$.

（1）若有 $s_i : s_j$ 不存在，根据[定理 4]即可得证.

（2）若 $s_i : s_j$ 都存在.若 $d$ 中序列只有两个，情形与[定理 4]同；若 $d$ 含有两个以上序列，由假设知必存在 $s_i, s_j, s_k \in d$，使得 $(s_i : s_j) : (s_i : s_k) \ne s_k : s_j$，取任意正整数三元组 $(p, q, r)$，参考[定理4]之证明即可.

若 $L$ 满足此必要条件，则称 $L$ 满足比例一致性.

**[定理6]** **$L_0$型序列集死锁判定定理**

给定 $L_0$ 型序列集 $L$ 及其展开式 $E = \{e_i \mid 1 \le i \le n\}$，顺序式关联划分 $D$.$L$ 无死锁当且仅当：

(1)$L$ 满足比例一致性.

(2)任意 $d = \{s_i \mid 1 \le i \le m\} \in D$ 及其最简周期表 $T_1 : T_2 : T_3 : ... : T_m$.$T_1 : T_2 : ... T_m = s_1.N : s_2.N : ... : s_m.N$ 或 $s_1.N = s_2.N = ... = s_m.N = \infty$，令 $L' = \cup prefix(e_i, T_i * s_i.Sequence.length)$，$L'$ 无死锁.

证明：(1)保证匹配的消息的数量上相等的可能性.(2)的前一个条件保证整个循环展开式有匹配的消息的数量上相等；后一个条件保证循环展开式周期片断上无死锁.然后根据[性质3]即可得证.

**[例2]** **$L_0$型序列集死锁分析举例[消息记法采用简化方式]**

表 2 循环模型死锁分析举例

| $s_1$ | $s_2$ | $s_3$ | 分析过程 |
|---|---|---|---|
|  |  |  |  |

| for(∞) | for(∞) | for(∞) | s₁ s₂ s₃彼此关联 |
|---|---|---|---|
| send(a) | recv(a), recv(c) | send(b), send(c) | 最简周期表$T_1:T_2:T_3$=6:3:2 |
| recv(b) | recv(c), send(a) | send(b), send(c) | 从$s_1$ $s_2$ $s_3$分别取前 6 次，3 次和 2 |
| end-for | end-for | send(c), send(b) | 次循环所获得的序列集死锁，故 |
|  |  | end-for | $s_1$ $s_2$ $s_3$ 死锁 |

## 5 条件模型死锁分析

### 5.1 条件模型的相关定义

#### [定义20] $C_0$型序列（集）和顺序式

序列 $s = Cs'$ 称为 $C_0$ 型序列，其中 $C$ 表示条件表达式，$s'$ 是条件辖域中的 $S_0$ 型序列，称为 $s$ 的顺序式，记为 $s_i.Sequence$ . $C_0$ 型序列的有限集合称为 $C_0$ 型序列集. $C_0$ 型序列集 $A = \{s_i \mid 1 \le i \le n\}$ 之顺序式 $A.Sequence$ 为 $\cup s_i.Sequence$ .

#### [定义21] $C_0$型序列（集）增量式

顺序式关联的 $C_0$ 型序列集 $L = \{s_i \mid 1 \le i \le n\}$，对每个 $s_j$，按如下过程构造 $s_j$ 增量式. $s_j$ 增量式记为 $s_j.Increase$ 或 $I(s_j)$ . $\{j, send, s_j.C, k\}$ 表示消息 $m(m.from = j, m.to = k)$，$\{j, recv, s_k.C, k\}$ 表示消息 $m(m.from = k, m.to = j)$ . $\cup I(s_i)$ 称为 $L$ 增量式，记为 $L.Increase$ 或 $I(L)$ .

```
for(s_j ∈ L)/*按s_j下标升序循环*/
    I(s_j) = ∅/*s_j增量式为S_0型序列，初始为空*/
    for(s_k ∈ {s | D_relation(s_j.Sequence, s_k.Sequence)})/*按s_k下标升序循环*/
        if(j < k) I(s_j) = I(s_j) + {j, send, s_j.C, k} + {j, recv, s_k.C, k}
        else I(s_j) = I(s_j) + {j, recv, s_k.C, k} + {j, send, s_j.C, k}
    end-for
end-for
```

#### [定义22] $C_0$型序列（集）替换式

顺序式关联的 $C_0$ 型序列集 $A = \{s_i \mid 1 \le i \le n\}$，$s_i$ 增量式与如下代码

if($I(s_j)$中所有收发的条件不是全部相等)　Debug: a deadlock will occur here

及 $s_i$ 构成 $s_i$ 替换式，记为 $s_i.Replace$ 或 $R(s_j)$ . $A$ 替换式为 $\cup s_i.Replace$，记为 $A.Replace$ 或 $R(A)$ .

### 5.2 条件模型死锁判定定理

#### [定理7] $C_0$型序列集增量式无死锁

$C_0$ 型序列集 $L$ 的增量式 $I(L) = \{I(s_i) \mid 1 \le i \le n\}$ 无死锁.

证明：(1)该增量式 $I(L)$ 初始为空. (2) $I(L)$ 中每个序列中的 $send$ 消息都是接收者的升序排

列，每个 $recv$ 消息都是发送者的升序排列.(3) $I(s_j) = I(s_j) + \{j, send, s_j.C, k\} + \{j, recv, s_k.C, k\}$ 与

$I(s_j) = I(s_j) + \{j, recv, s_k.C, k\} + \{j, send, s_j.C, k\}$ 执行次数相等.

当 $j = length(A)$，最后一次执行 $I(s_j) = I(s_j) + \{j, recv, s_k.C, k\} + \{j, send, s_j.C, k\}$ 之后，$E(L)$ 中宿主 $j$ 的最后两条信息依次为 $\{j, recv, s_k.C, k\}, \{j, send, s_j.C, k\}$.根据②可断定：此时 $I(L)$ 中宿主 $k$ 的序列的最后两条消息依次必定是 $\{k, send, s_k.C, j\}, \{k, recv, s_j.C, j\}$.根据匹配的定义，可知 $match(\{j, recv, s_k.C, k\}, \{k, send, s_k.C, j\})$ 且 $match(\{j, send, s_j.C, k\}, \{k, recv, s_j.C, j\})$.于是可令：

$I = I(L) -^R \{\{j, send, s_j.C, k\}, \{k, recv, s_j.C, j\}\} -^R \{\{j, recv, s_k.C, k\}, \{k, send, s_k.C, j\}\}$

类推且根据(1)(3)可得 $E$ 最终必定为空.根据[性质 2]可知 $I(L)$ 无死锁.

**[定理8]    $C_0$ 型序列集死锁必要条件**

$C_0$ 型序列集 $L$ 无死锁的必要条件是其增量式与顺序式皆无死锁.（证明略）

**[定理9]    $C_0$ 型序列集死锁充要条件**

$C_0$ 型序列集 $L$ 无死锁的充要条件是 $R(L)$ 无死锁.（证明略）

这个充要条件的主要意义是假如死锁，则给出死锁发生的位置.

**[例3]        $C_0$ 型序列集死锁分析举例[消息记法采用简化方式]**

表 3  条件模型死锁分析举例

| $s_1$ | $R(s_1)$ | $s_2$ | $R(s_2)$ | $s_3$ | $R(s_3)$ |
|---|---|---|---|---|---|
| if($c_1$) | send($c_1$,$s_2$) | if($c_2$) | recv($c_1$,$s_1$),send($c_2$,$s_1$) | if($c_3$) | recv($c_2$,$s_2$) |
| send(a) | recv($c_2$,$s_2$) | recv(b) | send($c_2$,$s_3$),recv($c_3$,$s_3$) | send(b) | send($c_3$,$s_2$) |
| | if($c_1 \neq c_2$) | recv(a) | if(!($c_1=c_2=c_3$)) | | if($c_2 \neq c_3$) |
| | deadlock | end-if | deadlock | | deadlock |
| | if($c_1$) | | if($c_2$) | | if($c_3$) |
| | send(a) | | recv(b),recv(a) | | send(b) |
| | | | end-if | | |
| 分析结果：编译前可判定 $s_1\ s_2\ s_3$ 顺序式，增量式无死锁；运行时给出可能死锁的位置 | | | | | |

## 6    结论与相关研究工作

本文给出了MPI同步通信程序三个重要的基本模型并给出了死锁判定定理.这些定理（以

及相关性质）既需编译前处理，部分也需运行时检测，因此是静态预防与动态检测的结合.有关 $c_0$ 型的定理和方法借鉴了文[19~20]中控制依赖转化为数据依赖的思想.有关 $L_0$ 型的定理和方法借鉴了专著[19]中取循环片断进行研究的思想. $S_0$ 型的定理可以证明文[11]中的死锁判定算法正确性（因为所有动态指令正好构成 $S_0$ 型），但需要注意他们的算法采用进程等待图（便于动态确定），我们采用的是消息依赖图（便于静态确定）.由于[11]需要在某条MPI调用之前插入握手代码，故容易使系统通信延迟迅速增加（比如对于长循环），文[12~14]则相当于将死锁监测代码封装到MPI库中，这也会大大增加系统开销.我们的方法仅仅在条件判定所辖MPI调用块（而非单个MPI调用）之前增加握手代码，这显然大大减少了死锁检测开销.与[15]（通道大小与数量等受限，参见原文第 3 章）等类似的工具和方法相比，我们的方法不受程序规模和大小的限制，也不必开辟多个进程(MPI节点数量通常都会超过[15]的分析能力)，但[15]可处理并发系统的诸多特性，而不限于死锁。

我们目前的模型还没有概括所有真实 MPI 程序（这正是下一步工作的重点），不过，这些基本模型和方法已奠定真实模型分析的必要基础.值得指出的是，这些方法完全可集成到[11~14]方法之中，加快死锁检测（比如在运行长循环之前就可判定是否死锁）.

**参考文献：**